\documentclass[doublecol,linenumbers]{epl2} 

\usepackage{graphicx}
\usepackage{amsmath}
\usepackage{bm}                         
\newcommand {\ba} {\begin{eqnarray}}
\newcommand {\ea} {\end{eqnarray}}

\title{The origin of the condensation energy scaling of Iron-based superconductors}
\shorttitle{The origin of the condensation energy scaling of Iron-based superconductors} 

\author{Yunkyu Bang}
\shortauthor{Yunkyu Bang}

\institute{Department of Physics, Chonnam National University, Kwangju 500-757, Republic of Korea
}

\abstract{The relation between the condensation energy (CE) and $T_c$ of a
phase transition reveals a fundamental nature of the transition.
In view of this, the recent experimental observation of the
non-BCS scaling relation of the CE {\it vs.} $T_c$ ($\Delta E \sim
T_c ^{3.5}$) with about forty different samples of the Fe-based
superconductors [J. Xing {\it et al.,} Phys. Rev. B 89, 140503
(2014)] was intriguing and strongly hinted at a non-BCS pairing
mechanism. In this paper, we have studied the CE and $T_c$ of the
multiband BCS model and found that the observed anomalous scaling
relation $\Delta E \sim T_c ^{3.5}$ is well reproduced by the
two-band BCS model paired by a dominant {\it repulsive interband
interaction} ($V_{inter} > V_{intra}>0 $). Our result implies that
this seemingly non-BCS-like scaling behavior of $\Delta E \sim T_c
^{3.5}$, on the contrary to the common expectations, is in fact a
strong experimental evidence that the pairing mechanism of the
Fe-based superconductors is genuinely a BCS mechanism, meaning
that {\it the Cooper pairs are formed by the itinerant carriers
glued by a pairing interaction.}}

\begin{document}

\maketitle

\section{Introduction}
The condensation energy (CE) $\Delta E$ of
a superconductor is defined as the energy difference between the
normal state and the superconducting (SC) state of the same
system. The size of the CE is a measure of how much stable the SC
state is, compared to the normal state, hence the CE is naturally
related to the transition temperature $T_c$. In more fundamental
context, the specific relation between the CE and the transition
temperature $T_c$ of a phase transition reveals the generic
characteristics of the phase transition. For example, the magnetic
transition with local moments such as a classic limit of Heisenberg model
and Ising model has the relation $\Delta E_{mag} \propto T_c$,
while the BCS theory of the one band
superconductor predicts $\Delta E_{BCS} \propto T_c ^2$\cite{BCS}.

These different power law relations are rooted to the fact that
the magnetic ordering above mentioned consists of local moments,
while the BCS SC ordering consists of itinerant electrons. In a
more revealing aspect, the former transition is determined solely
by the potential energy (PE) gain $\Delta PE <0$ by the ordering
of the local moments without kinetic energy involved. On the other
hand, the latter transition is determined by the subtle balance
between the kinetic energy (KE) loss $\Delta KE>0$ and the
potential energy gain $\Delta PE <0$ by the SC ordering.
Therefore, the standard scaling law $\Delta E \propto T_c ^2$ of
the BCS superconductor in fact reveals the fundamental nature of
the transition that the SC condensation is formed by itinerant
fermions.

In view of this observation, the recent report by J. Xing {\it et
al.}\cite{CE} of $\Delta E \propto T_c ^{\beta}$ ($\beta \approx
3.5$) with various Fe-based superconductors (IBS) is very intriguing and
should contain the crucial information of the SC pairing
mechanism. For example, the authors of \cite{CE} interpreted that
this anomalous scaling relation is a strong evidence that the
Fe-based superconductivity occur around the quantum criticality
(QC)\cite{Zaanen} and its pairing mechanism should be fundamentally
different from the BCS pairing mechanism. On the other hand,
regardless of the specific pairing mechanism, an extension of our
general discussion above leads us to speculate that {\it the
superconductivity of the IBS should have more
itinerant (kinetic) character than the standard one band BCS
superconductors.} We will show that this naive speculation is
indeed correct and we do not need to invoke the QC to understand
the anomalous scaling behavior of CE {\it vs.} $T_c$.

In this paper, we studied the CE and $T_c$ of a minimal two band
BCS model with various doping. To our surprise, this simple model immediately produced
the relation $\Delta E \propto T_c ^{3}$, which is a generic feature of the
two band BCS superconductor mediated by a dominant interband
pairing interaction (i.e. $|V_{inter}| > |V_{intra}|$). Then the
observed power $\beta \sim 3.5$ \cite{CE} can be easily obtained by adding
a small amount of pair-breaking impurities.

Although we cannot completely rule out other theoretical
explanations, our successful explanation of the
anomalous scaling relation of the CE {\it vs.} $T_c$ ($\Delta E \sim T_c ^{3.5}$\cite{CE}),
combined with the recent successful explanation\cite{SH_jump} of the another anomalous scaling
relation of the specific heat jump $\Delta C$ {\it vs.} $T_c$ (BNC scaling $\Delta C \sim
T_c^3$) \cite{BNC,BNC2,BNC3,JSKim,JSKim2,Hardy,Hardy2,Gofryk,Gofryk2}
using the same two band BCS model, has made a very strong case that {\it
the pairing mechanism of the IBS is genuinely
a BCS theory, with the multi-bands, mediated by a dominant inter-band pairing interaction.}
The strong correlation effects certainly should exist in these compounds in normal state, such as to
renormalize the quasiparticle masses $m_{qp}^{*}$ and the pairing
interactions $V_{ab}(\bf q)$, etc. However, the SC pairing mechanism itself seems to be
governed by the BCS theory.

\section{Condensation Energy}
\subsection{Two Band BCS model}
In real compounds of the IBS, there exist
multiple hole and multiple electron bands. In our minimal two band
model, each hole and electron band $\xi_{h(e)}(k)$, their DOS
$N_{h(e)}$, and the gap functions $\Delta_{h(e)}$ represent a
group of hole bands and a group of electron bands,
respectively\cite{SH_jump, Bang_model}. The Hamiltonian for the
two band superconductor is written as
\begin{eqnarray}
H &=& \sum_{k,\sigma} \xi_h(k) h^{\dag}_{k,\sigma} h_{k,\sigma} + \sum_{k,\sigma} \xi_e(k) e^{\dag}_{k,\sigma} e_{k,\sigma} \nonumber \\
&+&\sum_{k,k'} V_{hh} h^{\dag}_{k\uparrow} h^{\dag}_{-k\downarrow} h_{k'\downarrow} h_{-k'\uparrow}
+\sum_{k,k'} V_{ee} e^{\dag}_{k\uparrow} e^{\dag}_{-k\downarrow} e_{k'\downarrow} e_{-k'\uparrow} \nonumber \\
&+& \sum_{k,k'} V_{he} h^{\dag}_{k\uparrow} h^{\dag}_{-k\downarrow} e_{k'\downarrow} e_{-k'\uparrow}
+\sum_{k,k'} V_{eh} e^{\dag}_{k\uparrow} e^{\dag}_{-k\downarrow} h_{k'\downarrow} h_{-k'\uparrow} \nonumber \\
&&
\end{eqnarray}
\noindent where $V_{ab}=<V_{ab}(k,k')>_{FS}$ is Fermi surface (FS)
averaged pairing potentials with $a,b=h,e$ and assumed to have a
BCS cut-off, i.e. $|\xi_{h,e}(k)| < \Lambda_{hi}$ (which is
possibly the characteristic spin-fluctuation frequency,
$\omega_{sf}$, in Fe-based superconductors).

It is straightforward to extend the standard BCS theory to calculate the CE of the multi-band superconductor. Assuming the SC
order parameters(OPs) $\Delta_h$ and $\Delta_e$, the above
Hamiltonian is written in a quadratic form as follows,
\begin{eqnarray}
H &=& 2\sum_k \xi_h(k) h^{\dag}_k h_k + 2\sum_k \xi_e(k) e^{\dag}_k e_k \nonumber \\
&+&\sum_{k} \Delta_h h^{\dag}_k h^{\dag}_{-k} + \sum_{k} \Delta_h^{\ast} h_{k} h_{-k} \nonumber \\
&+&\sum_{k} \Delta_e e^{\dag}_k e^{\dag}_{-k} + \sum_{k} \Delta_e^{\ast} e_{k} e_{-k}
\end{eqnarray}
where $2$ is due to the spin degree of freedom (d.o.f.) and the self-consistent gap equations are defined as
\begin{eqnarray}
\Delta_h &=& -\sum_{k} V_{hh} b^h_k - \sum_{k} V_{he} b^e_k \nonumber \\
\Delta_e &=& -\sum_{k} V_{eh} b^h_k - \sum_{k} V_{ee} b^e_k
\end{eqnarray}
with the Cooper pair amplitudes $b^h_k = <h_{k\uparrow}
h_{-k\downarrow}>$ and $b^e_k = <e_{k\uparrow} e_{-k\downarrow}>$,
respectively. It is convenient to introduce the momentum summed
Cooper pair amplitudes $b^{h(e)}=\sum_k b^{h(e)}_k = \sum_k
\frac{1}{2}
\frac{\Delta_{h(e)}}{\sqrt{\xi_{h(e)}^2(k)+\Delta_{h(e)}^2}}=N_{h(e)}\Delta_{h(e)}
\log{\frac{\Lambda_{hi}+\sqrt{\Lambda_{hi}^2+\Delta_{h(e)}^2}}{|\Delta_{h(e)}|}}$ at $T=0$.
Unlike the single band gap equation, Eq.(3) shows that the
OPs $\Delta_{h(e)}$ and the Cooper pair amplitudes $b^{h(e)}$ are
coupled through $2 \times 2$ matrix $\hat{V}_{ab}$ such as
$\Delta_a = -\hat{V}_{ab} \cdot b^b$. Once we solve $\Delta_{h(e)}$
from the coupled gap equations Eq.(3), we numerically solve the inverse
matrix equation to obtain $b^{h(e)}$, with which $<H>_s$ of the
two-band Hamiltonian Eq.(2) is straightforwardly calculated. With
the general matrix equation $\Delta_a = -\hat{V}_{ab} \cdot b^b$,
there is no simple expression for $<H>_s$ as in one
band BCS superconductor\cite{Tinkham} except two special cases: (A)
$V_{he}=V_{eh}=0$, and (B) $V_{hh}=V_{ee}=0$.

The case (A) is a trivially decoupled two single band
superconductors with two independent $T_c$s and $\Delta_{h,e}$, respectively, for each band,
that is not our interest. The case (B) is more interesting case that is the pure
inter-band pairing limit and the analytic form of the CE $\Delta
E$ can be worked out as follows,
\begin{eqnarray}
\Delta E &=& <H>_s -<H>_n \nonumber \\
&=& \Delta KE + \Delta PE
\end{eqnarray}
with
\begin{eqnarray}
\Delta KE &=& \sum_{a=h,e} \sum_{k}\Bigl[|\xi_a(k)| - \frac{\xi_a ^2(k)}{E_a(k)}\Bigr]  \\
\Delta PE &=&  V_{he} b^{*h} b^e + V_{eh} b^{*e}b^{h} ,
\end{eqnarray}
\noindent where
$E_{h(e)}(k)=\sqrt{\xi_{h(e)}^2(k)+\Delta_{h(e)}^2}$. The kinetic
energy part $\Delta KE=\Delta KE_{h}+\Delta KE_{e}$ is calculated
in the same manner as in the single band model as\cite{Tinkham}
\begin{eqnarray}
\Delta KE_{h(e)} &=& -\frac{1}{2}N_{h(e)} \Delta_{h(e)}^2+\Big|\frac{\Delta_h \Delta_e}{V_{he}}\Big|
\end{eqnarray}
where we use the fact $V_{he}=V_{eh}$. The potential energy part
Eq.(6) is
\begin{equation} \Delta PE = \frac{\Delta_h^*
\Delta_e}{V_{he}} + \frac{\Delta_e^*
\Delta_h}{V_{eh}} = 2~ \frac{\Delta_h
\Delta_e}{V_{he}}
\end{equation}
\noindent where we assume $\Delta_{h(e)}^* =\Delta_{h(e)}$.
Therefore when $V_{he}>0$ (repulsive) and the OPs $\Delta_{h(e)}$
have the opposite signs each other, or when $V_{he}<0$
(attractive) and the OPs $\Delta_{h(e)}$ have the same signs each
other, we finally obtain the following simple expression
\begin{equation}
\Delta E = -\frac{1}{2}N_h \Delta_h ^2 - \frac{1}{2}N_e \Delta_e ^2 .
\end{equation}
The result of the total CE in Eq.(9) looks quite natural as the sum of the single band BCS CE of each band. However, this natural looking expression is totally
disguising because the $\Delta KE_{h(e)}$ and $\Delta PE_{h(e)}$ of each band
maximally depend on the other band with $V_{inter}\neq 0$ and $V_{intra}= 0$ (see Eq.(7) and (8)). As a consequence, the total CE $\Delta E$ of Eq.(9) will not follow the standard BCS scaling
form ($\Delta E_{BCS} \sim \Delta_0^2 \sim T_c^2$), but will produce a non-trivial
power-law relation because of the inverse relation $\frac{\Delta_h}{\Delta_e} \sim
\frac{N_e}{N_h}$ (at $T=0$) \cite{Bang_model} and the $T_c$-formula $T_c \sim \exp[-1/(V_{inter}\sqrt{N_hN_e})]$ \cite{SH_jump} of the two band BCS model mediated by the inter-band pairing interaction.

For general case when all $V_{ab}\ne 0 ~(a,b=h,e)$, the $\Delta
KE$  (Eq.(5)) can still be written as the following general
expression
\begin{equation}
\Delta KE = \sum_{a=h,e} -\frac{1}{2} N_a \Delta_a^2 + N_i \Delta_a^2
\log{\frac{\Lambda_{hi}+\sqrt{\Lambda_{hi}^2+\Delta_a^2}}{|\Delta_a|}},
\end{equation}
\noindent but the second term now cannot be reduced to the simple
expression $|\frac{\Delta_h\Delta_e}{V_{he}}|$ as in Eq.(7). Also the $\Delta PE$ now contains more terms than Eq.(6) as
\begin{equation}
\Delta PE =  V_{hh} b^{*h} b^h + V_{ee} b^{*e}b^{e}+V_{he} b^{*h} b^e + V_{eh} b^{*e}b^{h},
\end{equation}
\noindent and apparently $\Delta PE$ cannot have the simple
expression as Eq.(8). Therefore total CE $\Delta E$ and $T_c$ should be calculated numerically.

\subsection{Modeling of Doping}
To make a direct comparison with the experimental data of $\Delta
E$ {\it vs.} $T_c$ of the IBS\cite{CE}, which
are the collection of data of various iron Pnictides and
chalcogenide (FePn/Ch) SC compounds with systematic doping by
holes or electrons, we need a modeling of doping. Specifically,
we are modeling the cases of (Ba$_{1-x}$K$_x$)Fe$_2$As$_2$ and
Ba(Fe$_{1-x}$Co$_x$)$_2$As$_2$ which have the most systematic
doings in the experimental data.
We first note that the undoped parent compound BaFe$_2$As$_2$ is a
compensated metal, hence has the same number of electrons and
holes, i.e. $n_h = n_e$. Therefore it is a reasonable
approximation to assume $N_h \approx N_e$ at no doping and then
the doping of holes (K, Na, etc.) or electrons (Co, Ni, etc.) is
simulated by varying $N_h$ and $N_e$ while keeping $N_e + N_h
=N_{tot} = const.$ Admittedly this modeling of doping is much too
simple. However this assumption is qualitatively consistent with
the Angle-Resolved-Photo-Emission-Spectroscopy (ARPES)
measurements of (Ba$_{1-x}$K$_x$)Fe$_2$As$_2$\cite{Sato} and
Ba(Fe$_{1-x}$Co$_x$)$_2$As$_2$\cite{Sudayama} which show the
systematic changes of hole (electron) FS sizes with dopings.
Furthermore, the assumption $N_{tot} = const.$ is only for
convenience and can be relaxed. The sensitive parameter of our
model is the relative sizes between $N_e$ and $N_h$, but not the
total DOS $N_{tot}$. Finally, it is not necessary to know the
exact relation between the actual doping concentration $"x"$ of
real compounds and the values of $N_{h(e)}$ in our two-band model.
When we plot our calculation results of $\Delta E(N_{h(e)})$ {\it
vs.} $T_c (N_{h(e)})$, the explicit values of $N_{h(e)}$ become
hidden parameters and we only extract the scaling relation between
$\Delta E$ and $T_c$.

\section{Numerical results}

\begin{figure}[h]
\includegraphics[width=88mm]{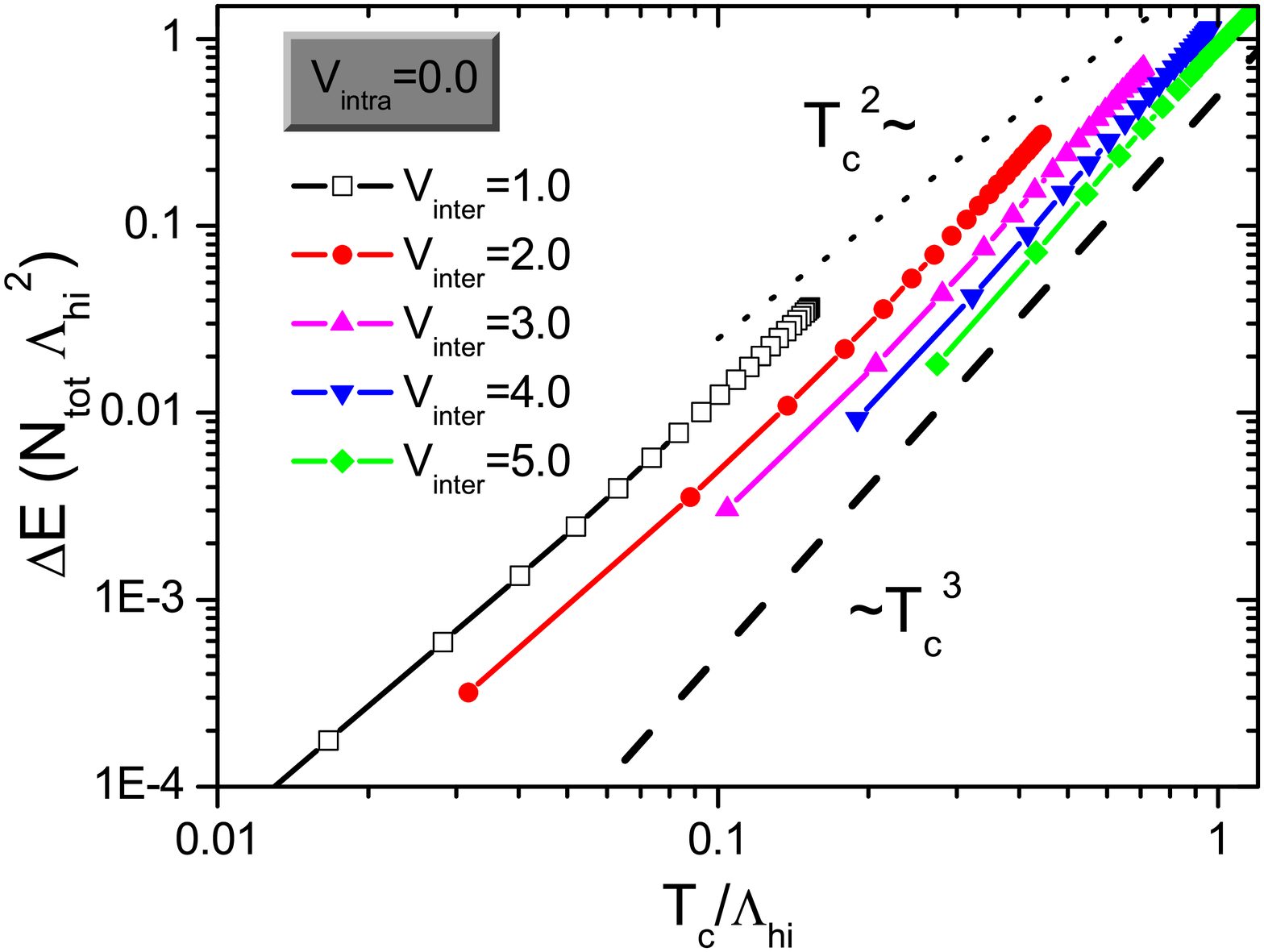}
\vspace{-0.5cm}
\caption{(Color online) Numerical calculations of $\Delta E$ {\it vs.} $T_c$ of the two band model with $\bar{V}_{intra}(=N_{tot}V_{intra})=0.0$
for $\bar{V}_{inter}(=N_{tot}V_{inter}) =1.0, 2.0, 3.0, 4.0$ and 5.0, respectively.
The dotted line ($\sim T_c^2$, BCS scaling) and the dashed line
($\sim T_c^3$) are guides for the eyes. \label{fig1}}
\vspace{1.4cm}
\includegraphics[width=88mm]{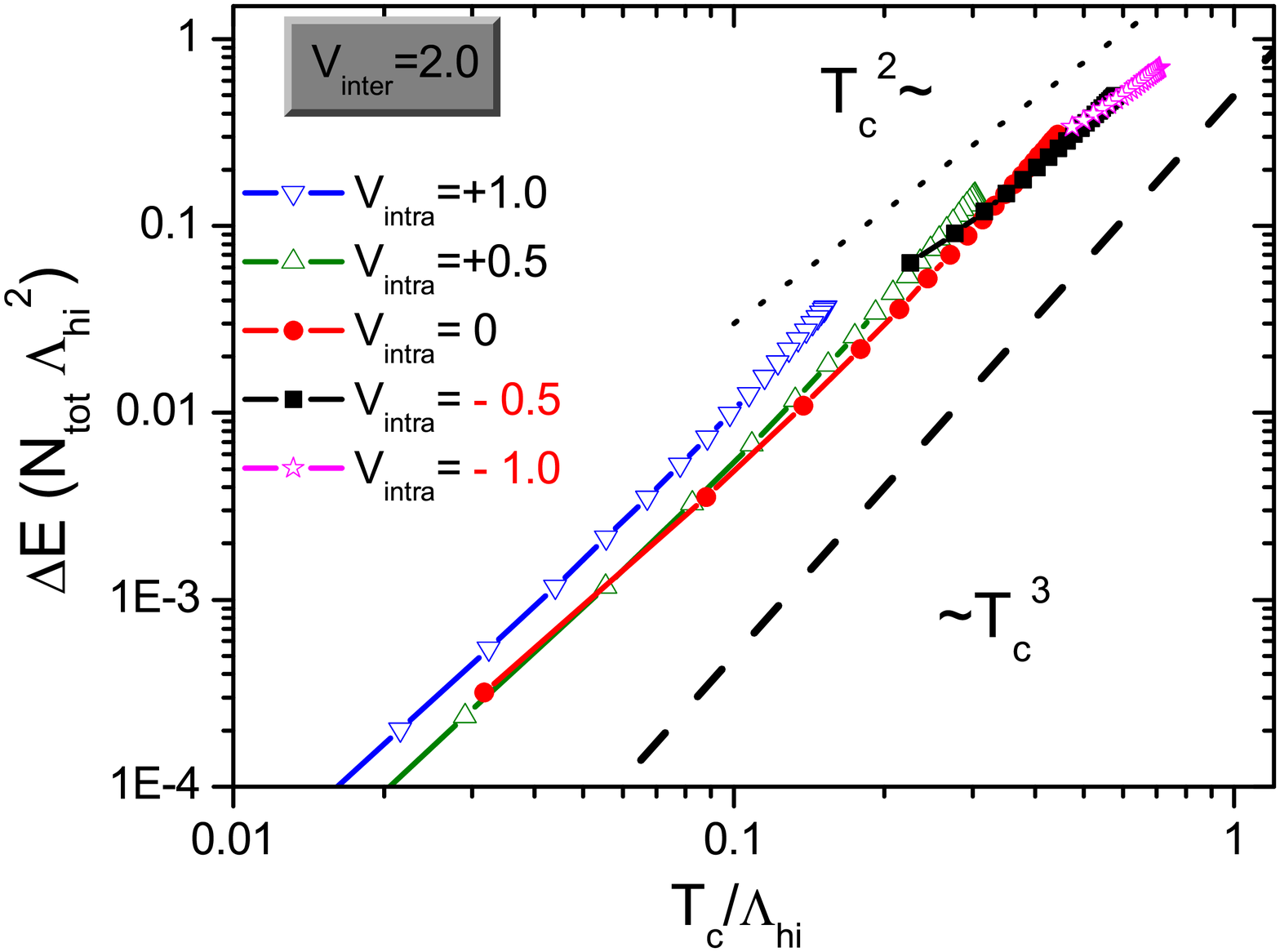}
\vspace{-0.5cm}
\caption{(Color online) Numerical calculations of $\Delta E$ {\it vs.} $T_c$ of the two band model with $\bar{V}_{inter}=2.0$
for $\bar{V}_{intra}=+1.0, +0.5, 0.0, -0.5$ and $-1.0$, respectively.
The dotted line ($\sim T_c^2$, BCS scaling) and the dashed line
($\sim T_c^3$) are guides for the eyes. \label{fig2}}
\end{figure}

\begin{figure}[h]
\noindent
\includegraphics[width=88mm]{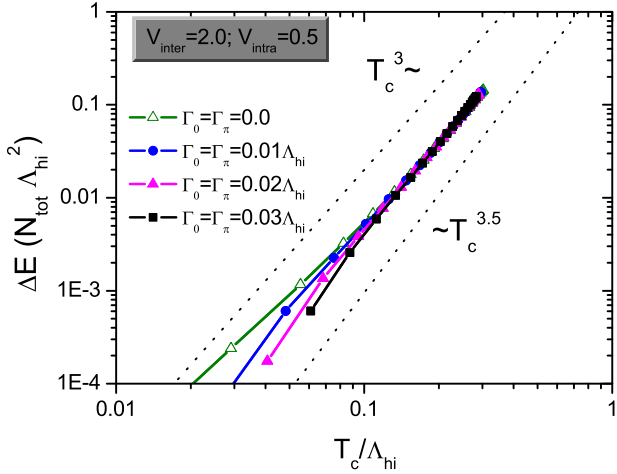}
\caption{(Color online) Numerical calculations of $\Delta E$ {\it vs.} $T_c$ of
the two band model with $\bar{V}_{inter}=2.0$ and $\bar{V}_{intra}=0.5$,
with the pair-breaking impurity scattering rates $\Gamma_{\pi}/\Lambda_{hi}=0.0, 0.01, 0.02,$ and 0.03, respectively.
The dotted lines of $\sim T_c^3$ and $\sim T_c^{3.5}$ are guides for the eyes. \label{fig3}}
\end{figure}

In all our numerical calculations, we only considered the
sign-changing s-wave ($s_{\pm}$) gap solutions assuming a
repulsive {\it inter-band} interaction ($V_{inter} >0$). Also
notice that all energy scales in this paper are normalized by
$N_{tot}$ such as $\bar{V}_{ab}=N_{tot}V_{ab}$ and
$\bar{N}_{h,e}=N_{h,e}/N_{tot}$. For given pairing interactions
$\bar{V}_{inter, intra}$, we solve the coupled gap equation Eq.(3)
with the mean field Hamiltonian Eq.(2) for continuously varying
$\bar{N}_h ~(\bar{N}_e=1-\bar{N}_h)$. In the $T \rightarrow 0$
limit, we obtain $\Delta_{h,e}$ and $b_{h,e}$ with which we can
calculate $\Delta E$ with Eqs.(10) and (11). $T_c$ can be calculated by taking
$\Delta_{h,e} \rightarrow 0$ limit. $\Delta E$ is negative
($<0$) by definition but we plot its absolute values in all
figures.

In Fig.1, we studied the case of the pure inter-band pairing case
(B): $\bar{V}_{intra}(=\bar{V}_{ee}=\bar{V}_{hh})=0$. We plotted
$\Delta E$ {\it vs.} $T_c$ for
$\bar{V}_{inter}(=\bar{V}_{he}=\bar{V}_{eh})=1.0, 2.0, 3.0, 4.0,$
and $5.0$, respectively.
To our surprise, this simple model
immediately produces a strongly non-BCS power law scaling $\Delta
E \sim T_c^{\beta}~ (\beta \approx 3)$ for a wide range of $T_c$, and
$\Delta E$ varies over three orders of magnitude. This is exactly
the key feature of the experimental data.
With the interaction strength $\bar{V}_{inter}$ used in Fig.1, the effective dimensionless coupling constant, $\lambda=\bar{V}_{inter}\sqrt{\bar{N}_h \bar{N}_e}$, runs from 0.3 to 2.5 ($\lambda_{min}=1.0\sqrt{0.1\times0.9}=0.3$ and  $\lambda_{max}=5.0\sqrt{0.5\times0.5}=2.5$). Therefore it is clear that the strongly non-BCS power law behavior of Fig.1 is independent of the weak or strong coupling limits and is not an artifact of the strong coupling limit.
Considering the extreme simplicity of two band model, we believe that this two-band
model  essentially captures the origin of the CE scaling behavior in
the IBS, that is, {\it a multi-band BCS
superconductor mediated by a dominant repulsive inter-band pairing
interaction $V_{inter}$.}

As discussed with Eq.(9), when $\bar{V}_{intra}=0$, the expression
of the total CE $\Delta E$ appears as a sum of two
single band BCS CE. Nevertheless, the result of Fig.1 shows that
it doesn't follow the standard BCS scaling law $\Delta E \sim T_c
^2$ with $N_{tot}=const.$, but follows much stronger power law
$\Delta E \sim T_c ^{\beta}$ with $\beta\approx 3$, even for a very weak coupling limit of $\bar{V}_{inter}=1.0$  ($0.3 < \lambda_{inter} <0.5$). As we
mentioned earlier, the origin of this strong power law is because
of the combined effect of the unique inverse relation of
$\frac{|\Delta_h|}{|\Delta_e|} \sim \frac{N_e}{N_h}$ and the
$T_c$-equation $T_c \sim \exp[-1/(V_{inter}\sqrt{N_hN_e})]$ of the
two band model in the pure inter-band pairing limit
\cite{SH_jump}. In fact, the results of Fig.1 can be exactly
reproduced with an {\it attractive} inter-band pairing
interaction ($\bar{V}_{inter} < 0$) with $s_{++}$-gap solution
($sign(\Delta_h)=sign(\Delta_e)$). However, we will find a
consistent evidence to choose $\bar{V}_{inter} > 0$ when we
consider the effect of the intra-band interaction
$\bar{V}_{intra}\neq 0$ in Fig.2.

In Fig.2, we studied the effect of the intra-band interaction
$\bar{V}_{intra} \neq 0$.  We experimented both repulsive
($\bar{V}_{intra}
> 0$) and attractive ($\bar{V}_{intra} < 0$) intra-band
interactions while fixing the inter-band interaction
$\bar{V}_{inter}=2.0$ at a moderate value; in this case the effective dimensionless coupling constant $\lambda_{inter}=\bar{V}_{inter}\sqrt{\bar{N}_h \bar{N}_e}$ varies for $0.6 < \lambda_{inter} < 1.0$.
First, when we add an attractive ($\bar{V}_{intra} < 0$) intra-band interaction, we
found: (1) $T_c$ quickly increases and the distribution of $T_c$s and $\Delta E$s
for varying $\bar{N}_h$ shrinks to the narrow range (see open pink
stars for $\bar{V}_{intra}=-1.0$ and black solid squares for
$\bar{V}_{intra}=-0.5$) of the upper right corner in Fig.2, which cannot be consistent with the three orders of magnitude variation of experimental data\cite{CE}.
(2) Also the scaling power quickly
converges to the BCS limit $\Delta E \sim T_c^2$ by even a very weak strength of the
attractive intra-band interaction $\bar{V}_{intra}=-0.5$ (black
solid squares). Again this cannot be compatible with experimental data.
On the other hand, adding a repulsive intra-band interaction ($\bar{V}_{intra} >0$),
the opposite trend occurs, i.e., the scaling relation becomes
steeper, in particular, more effective near the higher $T_c$
region. However, this enhancement of scaling power is still weaker to fit the experimental data
of $\Delta E \sim T_c^{3.5}$, and we need an addition mechanism to fit the data. Nevertheless,
the important  message of Fig.2 is that the intra-band interaction in the IBS should be
repulsive ($\bar{V}_{intra} > 0$) -- at least, not attractive.

Finally, in Fig.3, we studied the effect of the pair-breaking
impurity scattering (interband impurity scattering) $\Gamma_{\pi}$;
the non-pair-breaking impurity scattering (intraband impurity scattering) $\Gamma_{0}$
does not alter either $T_c$ or $\Delta_{h,e}$, therefore has no effect on the scaling behavior.
The physical non-magnetic impurities usually have both impurity potentials $\Gamma_{0}$ and $\Gamma_{\pi}$, and we considered in Fig.3 the maximum pair-breaking impurity, i.e. $\Gamma_{0} = \Gamma_{\pi}$\cite{Bang_imp}.
The results in Fig.3 show
that the scaling relation of $\Delta E$ {\it vs.} $T_c$ very quickly becomes steeper
with increasing the impurity scattering rate $\Gamma_{\pi}$. Even a very weak impurity
scattering rate $\Gamma_{\pi}/\Lambda_{hi}=0.03$ (black squares) is already
sufficient to make the scaling as $\Delta E \sim T_c^{3.5}$ as consistent with the experimental data\cite{CE}.
The technical reason why the impurity scattering $\Gamma_{\pi}$ increases the scaling power
as shown in Fig.3 is due to the fact that in general the pair-breaking impurity scattering
suppresses more efficiently the sizes of $\Delta_{h,e}$ than the size of $T_c$. The details of the impurity formalism is discussed in Appendix A.

\begin{figure}
\begin{center}
\includegraphics[width=55mm]{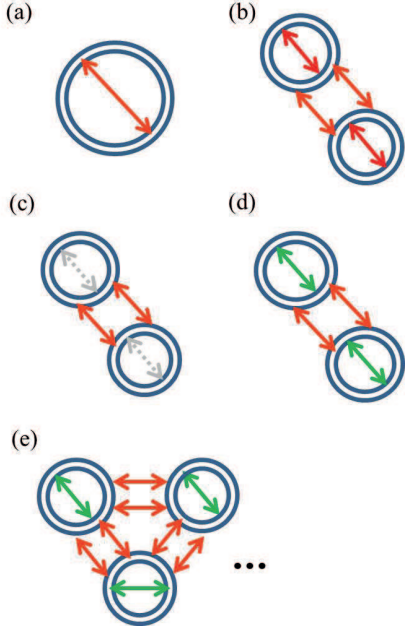}
\end{center}
\caption{(Color online)Illustrations of the KE losses and the PE gains
with the SC ordering for a single- and multi-band systems. Double lined circles represent the FSs
gapped by the SC ordering, hence the KE losses ($\Delta KE >0$) are all equal for all cases with
the same total DOS $N_{tot}$.
However, the PE gains ($\Delta PE <0$) differ in each case by the numbers of channels and the signs of the inter- and intra-band pairing interactions.
(a) A single band system with only an attractive (red arrow, $V_{intra}<0$) intra-band interaction
(standard BCS case). (b) Two band system with both attractive (red) inter-band ($V_{inter}<0$)
and intra-band interactions ($V_{intra}<0$); the total CE $\Delta E$ of this case is the same as (a).
(c) Two band system with an attractive inter-band interactions ($V_{inter}<0$) and
zero (grey dotted, $V_{intra}=0$) intra-band interactions. The intra-band part of the PE gain is lost
in this case as $\Delta PE_{intra}=0$.
(d) Two band system with an attractive (red)
inter-band interactions and repulsive (green) intra-band interactions. The PE gain is further reduced than (c) with $\Delta PE_{intra} > 0$.
(e) N-band system with $"N"$ repulsive (green) intra-band interactions and $"N(N-1)"$
attractive (red) inter-band interactions.\label{fig4}}
\end{figure}

\section{General Principle determining the scaling power $\beta$}
In Fig.1 and Fig.2, we have shown that the two band BCS model with dominant interband repulsion ($V_{inter} > V_{intra} >0$) displays the anomalous scaling relation $\Delta E \sim T_c^{\beta}$, $(\beta \approx 3)$ as a generic feature of the multi-band superconductor. We also found
that the repulsive (attractive) intra-band interactions $V_{intra}$ increases (decreases) the scaling power $\beta$. We would like to discuss this effect of the intra-band
interaction, $V_{intra} >0$ or $< 0$, on the scaling power $\beta$ in a more general context.
In the introduction, we have pointed out that the PE only transition such
as insulating magnets has $\Delta E \sim T_c$, while the BCS
superconductor which develops the transition by a subtle balance
between the $\Delta PE$ and $\Delta KE$ has $\Delta E \sim
T_c^2$. This balance between the $\Delta KE$ loss ($>0$) and the
$\Delta PE$ gain ($<0$) is the generic feature of the ordering
transitions of the itinerant fermion system such as
superconductivity and density wave (DW) transitions. In this view,
the higher power scaling relation $\Delta E \sim T_c^{3.5}$
observed in the IBS\cite{CE} implies that
the Fe-based superconductivity should be: either (1) suffering
more KE loss, and/or (2) not sufficiently harvesting the PE gain
through its transition. Both cases are not good news for us
wishing the maximum gain of CE in order to increase
$T_c$. We will show below that the cause (2) is the fundamental
origin of the steeper scaling relation ($\Delta E \sim T_c^{3.5}$)
of the IBS.

In Fig.4, we illustrate this energetic balance of the multi-band
SC systems. For a convenience of illustrations, we consider
{\it"attractive" inter-band} interaction ($V_{inter} <0$, red
color) with the $\Delta_{++}$ gap solution in Fig.4. Energetics-wise,
this is exactly the same as the {\it "repulsive" inter-band} interaction ($V_{inter} >0$) with the $\Delta_{\pm}$ gap solution.
With this assumption, we can systematically discuss the continuous evolution of energetics,
starting from the single band BCS superconductor to multi-band
systems.

Within the BCS theory, when FS opens a gap $\Delta_i$, the KE
increase (KE loss) in each band $"i"$ has always the same form as
\begin{equation}
\Delta KE_i = -\frac{1}{2} N_i \Delta_i^2 + N_i \Delta_i^2
\log{\frac{\Lambda_{hi}+\sqrt{\Lambda_{hi}^2+\Delta_i^2}}{|\Delta_i|}}
\end{equation}
\noindent where $N_i$ is the DOS and $\Delta_i$ is the gap of the
$i$th band. This form of $\Delta KE$ doesn't change with the
number of bands, and therefore assuming $N_{tot}=\sum_i N_i
=const.$, the total KE loss $\Delta KE$ should be approximately the
same size, depending on the average value of $\Delta_i$,
regardless of the number of bands. On the other hand, the PE gain
($\Delta PE <0$) of the multi-band systems differs by the numbers
of the channels and signs of the {\it inter-} and {\it intra-band} pairing
interactions, respectively.

Fig.4(a) is the case of a single band BCS superconductor with only
an attractive (red arrow, $V_{intra}<0$) intra-band interaction.
This system obtains the maximum PE gain
$\frac{\Delta^2}{V_{intra}}$. Combining the KE loss of Eq.(12),
the total CE becomes $\Delta E_{(a)} = -\frac{1}{2} N_0 \Delta^2$,
which is $\sim T_c^2$.
Fig.4(b) is the two band system with both attractive (red)
inter-band ($V_{inter}<0$) and intra-band interactions
($V_{intra}<0$). If the strengths of the interactions are the same
as $V_{inter}=V_{inter}$, this system is mathematically the same
as the single band system Fig.4(a). Hence we would get $\Delta
E_{(b)}\sim T_c^2$.
Fig.4(c) is the two band system with attractive (red) inter-band
interactions ($V_{inter}<0$) and zero (grey dotted, $V_{intra}=0$)
intra-band interactions. Compared to (b), the case (c) apparently
looses the "intra-part" of the PE gain ($\Delta PE_{intra}=0$).
And we found with the numerical calculations in Fig.1 that $\Delta
E_{(c)}\sim T_c^{\beta}$ with $\beta \approx 3$. Namely, the
scaling power $\beta$ has increased from the standard BCS value
$\beta_{BCS}=2$ because of the loss of the "intra-part" of the PE gain
compared to the case (b).

Fig.4(d) has a repulsive (green) intra-band
interactions ($V_{inter}>0$). Compared to the cases (b) and (c),
now $\Delta PE_{intra} (>0$) becomes not a gain but
a loss. Hence we can expect that the scaling power $\beta$ further
increase and that is indeed confirmed by the numerical
calculations in Fig.2. These case studies of (b), (c), and (d) tell
us that the intra-band interaction $V_{intra}$ in the real
compounds of the IBS cannot be attractive but
likely to be weakly repulsive as depicted in Fig.4(d).
And all our discussions of the KE losses and
the PE gains in Fig.4(b), (c), and (d) are unchanged if we replace
the {\it "attractive"} inter-band interaction $V_{inter} (<0$)
by the {\it "repulsive"} inter-band interaction $V_{inter} (>0$),
and the $\Delta_{++}$ gap -- assumed in this discussion -- by the $\Delta_{\pm}$ gap, so that now
Fig.4(d) has all repulsive interactions $V(\bf q)>0$, which is
consistent with the spin-fluctuation mediated interaction
scenarios\cite{Mazin,Mazin2,Hirschfeld}.

Finally, although physically not realistic but for a mathematical
completeness, we considered a $N$-band limit in Fig.4(e). In this
case, the number of repulsive intra-band interaction channels
increases as $\sim N$ and the number of attractive inter-band
interaction channels increases as $\sim N(N-1)$. In large-N limit,
the inter-band interaction channels dominate over the
intra-band channels and as a result the energetics of the system
converges back to the single band BCS limit; this N-band
system can be equivalently viewed as a N-sectioned single Fermi
pocket system. This analysis of a unphysical system, however, tells us
that the {\it two band system with a repulsive intra-band
interaction is the case with the maximum loss of
$\Delta PE_{intra}$, and therefore has the maximum scaling power
$\beta_{Max}\approx 3$}, unless other extrinsic effects such as a
pair-breaking impurity effect further increases it.

\section{Strong coupling effects}
Our study, being a mean field theory, didn't include the dynamical effects of the strong
coupling theory such as Eliashberg theory. In particular, Dolgov {\it et al.}\cite{Dolgov} have shown that the strong coupling effects, by inducing the mass-renormalization, qualitatively changes the relation between $\Delta_h /\Delta_e$ {\it vs.} $N_e /N_h$  from the weak coupling (BCS) theory that we used in this paper. However, this renormalization effect can be completely absorbed into our weak coupling formalism by replacing $N_{h(e)}$ in our theory with $\tilde{N}_{h(e)}=N_{h(e)}/[1+V_{inter}N_{h(e)}]$\cite{note3}.
Therefore, as far as the implicit parameter $R=\tilde{N}_{h}/\tilde{N}_{e}$ --although it is reduced compared to $R_0=N_{h}/N_{e}$ -- can vary in a substantial range of $[0,1]$, all the scaling results in this paper remain unchanged.
As a result, the basic energetics (the KE loss and the PE gain) of the multiband BCS
superconductor and its scaling relation of $\Delta E$ {\it vs.} $T_c$, should be generic and robust regardless of the weak or strong coupling limit.

\section{Summary and Conclusions}
We have studied the scaling relation of the $\Delta E$ {\it vs.}
$T_c$ of the two band BCS model. The doping (either holes or
electrons) in the real IBS compounds was modeled by continuously
varying the hole ($N_h$) and electron ($N_e$) DOSs keeping the
total DOS $N_{tot}=const.$ With numerical calculations, we found:
(1) $\Delta E \sim T_c^{\beta}$ with $\beta\approx 3$ is a generic
feature of the two band BCS superconductor with a dominant inter-band
pairing interaction (either attractive or repulsive).
(2) A repulsive intra-band interaction tends to increases the
scaling power $\beta$. On the other hand, an attractive intra-band
interaction, even if very weak, immediately turns the scaling
power into a BCS limit $\Delta E \sim T_c^2$. Therefore we are
forced to rule out a possible attractive intra-band interaction in
real compounds. (3) Adding a small amount of pair-breaking impurity
scattering ($\Gamma_{\pi}=0.05\Lambda_{hi}$) can easily increase
$\beta$ close to the experimental value of $\beta_{exp}\approx
3.5$.

We have also illustrated that the general principle determining
the scaling relation $\Delta E \sim T_c^{\beta}$ is the balance
between the KE loss and the PE gain through the SC transition. The origin of the
stronger scaling power of $\Delta E \sim T_c^{\beta}$
($\beta\approx 3$) than the standard BCS scaling of $\beta_{BCS}=2$
is due to the fact that the IBS, being a
multi-band BCS superconductor, didn't fully harvest all possible
PE gain, specifically, losing the intra-band PE part ($\Delta PE_{intra} >0$)
because of the repulsive $V_{intra}$.

In conclusion, our study implies that the
experimentally observed seemingly non-BCS scaling relation of
$\Delta E \sim T_c^{3.5}$\cite{CE} is in fact a strong
experimental evidence that the IBS are the
BCS superconductors, but with multi-bands, mediated by a dominant
repulsive inter-band pairing interaction. All strong correlation
effects, abundantly observed in the normal state, should
renormalize the effective mass $m_{qp}^{*}$ of quasiparticles, DOS $N_{h,e}$,
pairing interactions $V_{inter,intra}(\bf q)$, etc., but when
the system enters the SC transition, the pairing mechanism itself
seems to be governed by the BCS mechanism.
Finally, the results of this paper has no direct relevance to the FeSe monolayer
and related systems where only electron FSs exist\cite{FeSe}.

\acknowledgments
This work was supported by Chonnam
National University Grant 2014 and NRF Grant 2013-R1A1A2-057535
funded by the National Research Foundation of Korea. The author
gratefully acknowledge insightful discussions with G. R. Stewart.

\section{Appendix A: Impurity scattering formalism}
In this paper, we only consider the non-magnetic impurities, and for our two band BCS model,
the non-magnetic impurity scattering process can be
conveniently described by two impurity scattering rate parameters:
$\Gamma_0$ (intra-band) and $\Gamma_{\pi}$ (intra-band).
And all impurity scattering effects enter the coupled gap equations Eq.(3) in the main text through
the pair susceptibilities $\chi_{h,e}$, which is defined as follows.
\begin{equation}\tag{A1}
b^{h(e)} = \sum_k b^{h(e)}_k = \Delta_{h(e)} \chi_{h,e}(T),
\end{equation}
with $b^h_k = <h_{k\uparrow}
h_{-k\downarrow}>$ and $b^e_k = <e_{k\uparrow} e_{-k\downarrow}>$,
respectively. Without impurities, these pair susceptibilities $\chi_{h,e}(T)$ can
be analytically calculated in BCS limit,
but with impurities they need numerical calculations with the following expression,

\begin{equation}\tag{A2}
\chi_{h,e}(T) = T\sum_n N_{h,e}\int_{-\Lambda_{sf}}^{\Lambda_{sf}} d\xi \frac{\delta_{h,e}}{\tilde{\omega}_n^2 + \xi^2 + \tilde{\Delta}^2_{h,e}}.
\end{equation}
where $\tilde{\omega}_n=\omega_n+(\Gamma_0+\Gamma_{\pi})$ is the renormalized Matsubara frequency by the non-magnetic impurity scattering, and conveniently parameterized as $\tilde{\omega}_n=\omega_n \eta_0$ with $\eta_0=1+(\Gamma_0+\Gamma_{\pi})/|\omega_n|$. $\delta_{h,e}$ are the renormalization parameter of the OPs $\Delta_{h,e}$ due to impurities, which is defined as $\delta_{h,e}=1+(\Gamma_0 + a_{h,e}\Gamma_{\pi})/|\omega_n|$, with $a_{h}=\frac{(N_e \Delta_{e})}{(N_h \Delta_{h})}$  and $a_{e}=\frac{(N_h \Delta_{h})}{(N_e \Delta_{e})}$; $a_{h,e}$ are always negative number, because of the opposite signs of $\Delta_{h}$ and $\Delta_{e}$, and its size is $\sim O(1)$\cite{Bang_imp}; and if $\Gamma_{0}=\Gamma_{\pi}$, $\delta_{h,e} \approx 1$.
We can clearly see that $\Gamma_0$ alone does not affect $T_c$ and $\Delta_{h,e}$,
because when $\Gamma_{\pi}=0$ and $\Gamma_{0}\neq 0$, then $\eta_0 = \delta_{h,e}$, and in this case the pair susceptibilities $\chi_{h,e}(T)$ are not renormalized with impurities\cite{Anderson_imp,AG}; therefore, only $\Gamma_{\pi}$ is pair-breaking\cite{Bang_imp}. In this paper, we considered the maximum pair-breaking non-magnetic impurities, i.e., $\Gamma_{0}=\Gamma_{\pi}$, and in this case $\delta_{h,e} \approx 1$.

Regarding the systematic enhancement of the scaling power in $\Delta E \sim T_c^{\beta}$ with impurity scattering $\Gamma_{0}=\Gamma_{\pi} \neq 0$ in Fig.3 in the main text, we argued that its origin is because the impurity suppression rate of OPs $\tilde{\Delta}_{h,e}$ is faster than the impurity suppression rate of $T_c$. To see this, the $T_c$-suppression is determined by $\chi_{h,e}(T=T_c)$,
\begin{equation}\tag{A3}
\chi_{h,e}(T_c) \approx T_c\sum_n N_{h,e}\int_{-\Lambda_{sf}}^{\Lambda_{sf}} d\xi \frac{1}{(\omega_n \eta_0)^2 + \xi^2}.
\end{equation}
With the above renormalized susceptibility, we can read that $T_c$ is reduced in linear with $\Gamma_{\pi}$ as $T_c \approx T_c^0 -\frac{\pi}{4}\Gamma_{\pi}$\cite{AG}. On the other hand, the $\Delta_{h,e}$-suppression is determined by $\chi_{h,e}(T\rightarrow 0)$ as

\begin{equation}\tag{A4}
\chi_{h,e}(T\rightarrow 0) \approx T\sum_n N_{h,e}\int_{-\Lambda_{sf}}^{\Lambda_{sf}} d\xi \frac{1}{(\omega_n \eta_0)^2 + \xi^2 + \tilde{\Delta}^2_{h,e}}.
\end{equation}
When the above susceptibilities are substituted into the gap equations Eq.(3) in the main text, it is clear that the renormalized $\tilde{\Delta}_{h,e}$ are determined by the non-linear equations, and this non-linear suppression of $\tilde{\Delta}_{h,e}$ is systematically enhanced as $\tilde{\Delta}_{h,e}$ become smaller values, as seen in the lower-left corner region in Fig.3.
As results of all these effects, Fig.3 shows that the low $T_c$ tail
of the $\Delta E$ {\it vs.} $T_c$ scaling becomes systematically steeper with impurity scattering.


\begin{thebibliography}{0}
\bibitem{note3}
These redefined DOSs $\tilde{N}_{h(e)}=N_{h(e)}/[1+V_{inter}N_{h(e)}]$ are not the commonly defined (physical) renormalized DOSs, which are rather $\tilde{N}^{phy}_{h(e)}=N_{h(e)} \cdot [1+V_{inter}N_{e(h)}]$.
\end{thebibliography}
\end{document}